%% file: ms.tex
\documentclass[]{elsarticle}

\makeatletter
\def\ps@pprintTitle{%
 \let\@oddhead\@empty
 \let\@evenhead\@empty
 \def\@oddfoot{}%
 \let\@evenfoot\@oddfoot}
\makeatother

\usepackage{physics}
\usepackage{natbib}   % Literaturverzeichnis
\usepackage[unicode]{hyperref}
\usepackage{cleveref}
\usepackage{caption}
\usepackage{subcaption}
\usepackage{pgfplots}
\pgfplotsset{compat=1.18}
\graphicspath{{figures/}}
\usepackage{geometry}
\usepackage{todonotes}

\begin{document}

\begin{frontmatter}

\title{Analytical and numerical validation of a plate-plate tribometer for measuring wall slip}

\author[]{Muhammad Hassan Asghar$^{1}$}
\ead{asghar@mma.tu-darmstadt.de}

\author[]{{Tomislav Mari\'{c}}\corref{corr}$^{1}$}
\cortext[corr]{Corresponding author}
\ead{maric@mma.tu-darmstadt.de}

\author[]{Houssem Ben Gozlen$^{2}$}
\ead{bengozlen@fdy.tu-darmstadt.de}

\author[]{Suraj Raju$^{1}$}
\ead{raju@mma.tu-darmstadt.de}

\author[]{Mathis Fricke$^{1}$}
\ead{fricke@mma.tu-darmstadt.de}

\author[]{Maximilian Kuhr$^{3}$}
\ead{maximilian.kuhr@tu-darmstadt.de}

\author[]{Peter F. Pelz$^{3}$}
\ead{peter.pelz@tu-darmstadt.de}

\author[]{Dieter Bothe$^{1}$}
\ead{bothe@mma.tu-darmstadt.de}

\address{$^{1}$Mathematical Modeling and Analysis Group, TU Darmstadt \\ 
$^{2}$Chair of Fluid Dynamics, TU Darmstadt\\
$^{3}$Chair of Fluid Systems, TU Darmstadt
}

\begin{abstract}

We model the Darmstadt Slip Length Tribometer (SLT) originally presented by \citet{dglt}. The plate tribometer is specially designed to measure viscosity and slip length simultaneously for lubrication gaps in the range of approximately 10 micrometres at relevant temperatures and surface roughness. We investigate the inlet effect of the flow on the results by varying the inner radius of the fluid inlet pipe. The outcomes of numerical simulations suggest that variations in the diameter of this inner radius have minimal impact on the results. Specifically, any alterations in the velocity profile near the inlet, brought about by changes in the diameter, quickly revert to the profile predicted by the analytical model. The main conclusion drawn from this study is the validation of the Navier-Slip boundary condition as an effective model for technical surface roughness in CFD simulations and the negligible influence of the inlet effect on the fluid dynamics between the tribometer's plates.

\end{abstract}

% Keywords
\begin{keyword}
    lubrication theory, Navier slip length, tribometer, finite volume method, simulation
\end{keyword}

\end{frontmatter}

%\linenumbers

\input{sections/introduction}
\input{sections/methods-and-models}
\input{sections/results}
\clearpage
\input{sections/conculsion}

%\clearpage
\section{Acknowledgements}

We acknowledge the financial support by the German Research Foundation (DFG) within the Collaborative Research Centre 1194 (Project-ID 265191195).

%The use of the high-performance computing resources of the Lichtenberg High-Performance Cluster at the TU Darmstadt is gratefully acknowledged.

\clearpage

\bibliography{bibliography}

\end{document}

%% file: sections/introduction.tex
\section{Introduction}

For many flow problems in science or engineering, the no-slip boundary condition is applied to model the interaction between the fluid flow and the surrounding solid walls. The no-slip condition states that the fluid velocity at an impermeable wall is identical to the velocity of the wall itself, preventing relative motion between the wall and the fluid molecules at the wall. Even if the no-slip boundary condition is an appropriate model for many technical applications, However, deviations have been observed for microfluidic flows or dynamic wetting flows, where the breakdown of the no-slip boundary condition has been clearly pointed out in \cite{Huh1971}. This insight is by no means new, as the slip boundary condition was already postulated by Claude-Louis Navier~\citep{navier1822memoire} in 1822. He postulated that the fluid slips, with a relative velocity linearly proportional to the shear rate at the wall. Here, the constant of proportionality is the slip length, which, according to \citet{Helmholtz.1860}, can be interpreted as an effective increase in gap clearance. The Navier slip boundary condition expresses a balance between the friction force (opposing the tangential motion) and the component of the viscous shear force parallel to the solid wall, i.e., \
\begin{align}\label{eqn:navier_slip_1}
-\beta \left( \boldsymbol{v} - \boldsymbol{v}_w \right)_\parallel = \mathbf{(Sn)}_{||}.
\end{align}
Here $\beta$ is a coefficient describing the amount of friction between liquid and solid particles, $\mathbf{S} = \mu (\nabla \boldsymbol{v} + \nabla \boldsymbol{v} ^T)$ is the viscous stress tensor and $\mu$ denotes the dynamic viscosity. Within continuum thermodynamics, equation \ref{eqn:navier_slip_1} is the simplest, namely linear, closure to model the a priori unknown relative tangential velocity which appears in the entropy production contribution due to relative (tangential) motion between two phases; see \cite{Bothe2022sharp} for more details. Moreover, the solid wall is assumed to be impermeable, which implies the normal component of the relative velocity vanishes, i.e.\
\begin{equation}\label{eqn:imp_cond}
    \left( \boldsymbol{v} - \boldsymbol{v}_w \right) \cdot \boldsymbol{n} = 0,
\end{equation}
where $\boldsymbol{n}$ denotes the outer normal field to the solid wall.
Note that the full boundary condition for solving the Navier-Stokes (or Stokes) equations for fluid flow in a confined domain with Navier-slip is given by (1) and (2) together.  

An equivalent formulation of \eqref{eqn:navier_slip_1} is obtained by dividing the equation by the friction coefficient, leading to
\begin{align}\label{eqn:navier_slip_2}
\left( \boldsymbol{v} - \boldsymbol{v}_w \right)_\parallel + 2 \lambda \mathbf{(Dn)}_{||} = 0.
\end{align}
Here, $\mathbf{D}=\frac12 (\boldsymbol{v} + (\nabla \boldsymbol{v}) ^T)$ is the rate-of-deformation tensor and the quantity
\begin{align}
\lambda := \frac{\mu}{\beta} 
\end{align}
is called the \emph{slip length}, where $\mu$ denotes the dynamic viscosity of the fluid. It is known from molecular dynamics simulations and experimental investigation that the value of the slip length is typically of the order of nanometers \cite{Neto2005}. Hence, it is far below the characteristic length scales of the flow for many technical applications. Furthermore, the impact of $\lambda$ on the macroscopic flow may be small, particularly for single-phase flows. However, for dynamic wetting flows or flow processes where the characteristic length scale is of the order of the slip length, e.g., sealing systems and hydraulic components, it is important to determine the value of the slip parameter. For this reason, the \emph{Darmstädter Slip length Tribometer} (SLT) was developed~\citep{dglt}. The SLT is a classic plate tribometer that measures the frictional torque transmitted from the rotating plate through the entrapped liquid to the stationary plate. An analytical solution of Navier-Stokes equations with Navier-slip boundary condition for a flow between two rotating plates then relates the slip length to the measured torque. The aim of the present work is to show, utilizing CFD simulations in the open-source software OpenFOAM~\citep{openfoam1,Maric2014}, that the simplified analytical model 
\begin{equation}\label{eqn:torque-formula}
    M = \frac{\mu I_p \Omega}{h+2\lambda}
\end{equation}
for the torque $M$ as a function of the plate distance $h$ and angular velocity $\Omega$ is sufficient to infer the slip length $\lambda$ from the experimentally measured torque. This conclusion is not obvious because the expression \eqref{eqn:torque-formula} can only be derived from a solution of the Stokes equations that disregards the feed flow entering and leaving the device during the measurement.

% \\
% \[ *** \]
% \textbf{TODO:} Shall we still cite \citep{DissCorneli} and \citep{ArticleCorneli}? Please put the citation in an appropriate position.
%%%%%%%%%%%%%%%%%%%%%%%%%%%%%%%%%%%%%%%%%%

%% file: sections/methods-and-models.tex
\section{Materials and Methods}
\label{sec:methodsAndModels}
%%%%%%%%%%%%%%%%%%%%%%%%%%%%%%%%%%%%%%%%%%%%%%%%%%%%%%
\subsection{Analytical model}
\begin{figure}[ht!]
    \def\svgwidth{0.5\textwidth}
    {\footnotesize
        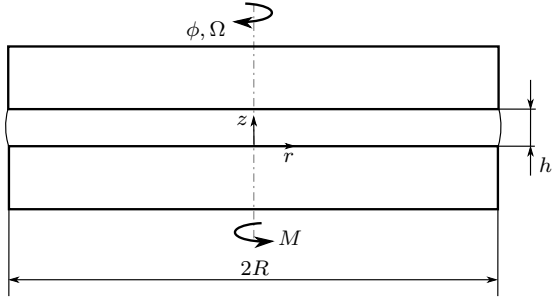
    }
    \centering
    \caption{Principle sketch of the analytical model \cite{DissCorneli}.}   
    \label{Sketch_Model1}
\end{figure}

We consider the steady and rotationally symmetric flow of an incompressible Newtonian fluid between two plates of radius R. The upper plate rotates with constant angular velocity $\Omega$. The lower plate is at rest. Let  $(r, \phi, z)$ be the cylindrical coordinates so that the plates are at $z = 0$ and $z = h$. Figure \ref{Sketch_Model1} shows a sketch of the tribometer.
\subsubsection{Equations of motion}

Compared to \citep{DissCorneli}, where a simpler linear model is assumed, we apply Navier-slip boundary conditions to full steady-state Navier-Stokes equations in cylindrical coordinates.

We introduce the velocity vector $\boldsymbol{v}=v_r \boldsymbol{e_r}+v_{\phi} \boldsymbol{e_{\phi}}+v_z \boldsymbol{e_z}$ and the pressure $p$. Considering the rotational symmetry, the Navier-Stokes equations have the following form \cite{Spurk}
\begin{equation}
    \pdv{v_r}{r} + \pdv{v_z}{z} + \frac{v_r}{r} = 0,
    \label{eq:d_nv1}
\end{equation}
\begin{equation}
    v_r \pdv{v_r}{r} + v_z \pdv{v_r}{z} - \frac{v_{\phi}^2}{r}
                     = -\frac{1}{\rho} \pdv{p}{r} +\nu \left(\pdv[2]{v_r}{r} + \frac{1}{r}\pdv{v_r}{r} - \frac{v_r}{r^2} + \pdv[2]{v_r}{z} \right),
    \label{eq:d_nv2}
\end{equation}
\begin{equation}
    v_r \pdv{v_{\phi}}{r} + v_z \pdv{v_{\phi}}{z} + \frac{v_r v_{\phi}}{r}
                    =\nu \left(\pdv[2]{v_{\phi}}{r} + \frac{1}{r}\pdv{v_{\phi}}{r} - \frac{v_{\phi}}{r^2} + \pdv[2]{v_{\phi}}{z} \right),
    \label{eq:d_nv3}
\end{equation}
\begin{equation}
     v_r \pdv{v_z}{r} + v_z \pdv{v_z}{z}
                    = -\frac{1}{\rho} \pdv{p}{z} + \nu \left(\pdv[2]{v_z}{r} + \frac{1}{r}\pdv{v_z}{r} + \pdv[2]{v_z}{z} \right).
    \label{eq:d_nv4}
\end{equation}
Here, $\rho$ denotes the density, and $\nu$ is the kinematic viscosity. 

\subsubsection{Non-dimensionalization of the Equation of Motion}
The plate radius $R$ and the gap width $h$ are the characteristic lengths to non-dimensionalize the equations. We define the dimensionless variables
\begin{equation}
    \hat{r} :=\frac{r}{R}, \quad \hat{z} :=\frac{z}{h}, \quad \hat{v}_r :=\frac{v_r}{R \Omega}, \quad \hat{v}_{\phi} :=\frac{v_{\phi}}{R \Omega}, \quad \hat{v}_z:=\frac{v_z}{h \Omega}, \quad \hat{p}:=\frac{p}{\rho R^2 \Omega^2}.
\end{equation}
Substituting the dimensionless variables into the dimensioned \cref{eq:d_nv1,eq:d_nv2,eq:d_nv3,eq:d_nv4} yields the dimensionless system of equations as
\begin{equation}
    \pdv{\hat{v}_r}{\hat{r}} + \pdv{\hat{v}_z}{\hat{z}} + \frac{\hat{v}_r}{\hat{r}} = 0,
    \label{eq:nd_nv1}
\end{equation}
\begin{equation}
    \hat{v}_r \pdv{\hat{v}_r}{\hat{r}} + \hat{v}_z \pdv{\hat{v}_z}{\hat{z}} - \frac{\hat{v}_{\phi}^2}{\hat{r}}
        = - \pdv{\hat{p}}{\hat{r}} + \frac{1}{Re} \left[ \delta^2 \left( \pdv[2]{\hat{v}_r}{\hat{r}} + \frac{1}{\hat{r}}\pdv{\hat{v}_r}{\hat{r}} - \frac{\hat{v}_r}{\hat{r}^2} \right) 
          + \pdv[2]{\hat{v}_r}{\hat{z}} \right],
    \label{eq:nd_nv2}
\end{equation}
\begin{equation}
    \hat{v}_r \pdv{\hat{v}_{\phi}}{\hat{r}} + \hat{v}_z \pdv{\hat{v}_{\phi}}{\hat{z}} + \frac{\hat{v}_r \hat{v}_{\phi}}{\hat{r}} 
        = \frac{1}{Re} \left[ \delta^2 \left( \pdv[2]{\hat{v}_{\phi}}{\hat{r}} + \frac{1}{\hat{r}}\pdv{\hat{v}_{\phi}}{\hat{r}} - \frac{\hat{v}_{\phi}}{\hat{r}^2} \right)
         + \pdv[2]{\hat{v}_{\phi}}{\hat{z}} \right],
    \label{eq:nd_nv3}
\end{equation}
\begin{equation}
    \hat{v}_r \pdv{\hat{v}_z}{\hat{r}} + \hat{v}_z \pdv{\hat{v}_z}{\hat{z}}
    = - \frac{1}{\delta^2} \pdv{\hat{p}}{\hat{z}} + \frac{1}{Re} \left[ \delta^2 \left( \pdv[2]{\hat{v}_z}{\hat{r}} + \frac{1}{r}\pdv{\hat{v}_z}{\hat{r}}\right) 
      + \pdv[2]{\hat{v}_z}{\hat{z}} \right].
    \label{eq:nd_nv4}
\end{equation}
The non-dimensional numbers used in this study are the Reynolds number $Re=\left({h^2 \Omega}\right)/{\nu}$ and the plate spacing $\delta={h}/{R}$. 
According to the operation parameters of the SLT, the Reynolds number is of the order of $10^{-9}$. Therefore, the inertial terms can be neglected, and the following Stokes equations apply
\begin{equation}
    \pdv{\hat{v}_r}{\hat{r}} + \pdv{\hat{v}_z}{\hat{z}} + \frac{\hat{v}_r}{\hat{r}} = 0,
\end{equation}
\begin{equation}
    - \pdv{\hat{p}}{\hat{r}} + \frac{1}{Re} \left[ \delta^2 \left( \pdv[2]{\hat{v}_r}{\hat{r}} + \frac{1}{\hat{r}}\pdv{\hat{v}_r}{\hat{r}} - \frac{\hat{v}_r}{\hat{r}^2} \right)
        + \pdv[2]{\hat{v}_r}{\hat{z}} \right] = 0,
\end{equation}
\begin{equation}
    \frac{1}{Re} \left[ \delta^2 \left( \pdv[2]{\hat{v}_{\phi}}{\hat{r}}+\frac{1}{\hat{r}}\pdv{\hat{v}_{\phi}}{\hat{r}} - \frac{\hat{v}_{\phi}}{\hat{r}^2} \right)
     + \pdv[2]{\hat{v}_{\phi}}{\hat{z}} \right] = 0,
    \label{NS_kap3_phi}
\end{equation}
\begin{equation}
    - \frac{1}{\delta^2} \pdv{\hat{p}}{\hat{z}} + \frac{1}{Re} \left[ \delta^2 \left( \pdv[2]{\hat{v}_z}{\hat{r}} + \frac{1}{\hat{r}}\pdv{\hat{v}_z}{\hat{r}}\right)
        + \pdv[2]{\hat{v}_z}{\hat{z}} \right] = 0.
\end{equation}
It should be emphasized that the equation \eqref{NS_kap3_phi} is uncoupled and can be solved independently. Since we are only interested in the torque at the bottom plate resulting from the shear stress $\tau_{\phi z}=\mu \pdv{v_{\phi}}{z}$, only the circumferential velocity component $\hat{v}_\phi(\hat{r},\hat{z})$ will be computed. For simplicity, the circumferential velocity component $\hat{v}_\phi(\hat{r},\hat{z})$ is denoted as $\hat{v}(\hat{r},\hat{z})$. The equation 
\begin{equation}
    \delta^2 \left( \pdv[2]{\hat{v}}{\hat{r}} + \frac{1}{\hat{r}}\pdv{\hat{v}}{\hat{r}} - \frac{\hat{v}}{\hat{r}^2} \right)+\pdv[2]{\hat{v}}{\hat{z}} = 0
    \label{eqn_Bew_Gl}
\end{equation}
is solved in the following.

%%%%%%%%%%%%%%%%%%%%%%%%%%%%%%%%%%%%%%%%%%%%%%%%%%%%%

\subsubsection{Boundary conditions}

In order to solve the equation of motion \eqref{eqn_Bew_Gl} uniquely, the boundary conditions must be specified. The Navier slip boundary condition is applied at the plates, and considering the rotational symmetry, the following boundary conditions apply
\begin{equation}
    \hat{v}(\hat{r} = 0,\hat{z}) = 0,
    \label{RB1}
\end{equation}
\begin{equation}
    - \frac{\lambda_u}{h} \pdv{\hat{v}(\hat{r},\hat{z})}{\hat{z}} \bigg|_{\hat{z}=0} + \hat{v}(\hat{r},\hat{z} = 0) = 0,
    \label{RB2_v1}
\end{equation}
and
\begin{equation}\label{RB3_v1}
      \frac{\lambda_o}{h} \pdv{\hat{v}(\hat{r},\hat{z})}{\hat{z}} \bigg|_{\hat{z}=1} + \hat{v}(\hat{r},\hat{z}=1)=\hat{r}.
\end{equation}
%

%%%%%%%%%%%%%%%%%%%%%%%%%%%%%%%%%%%%%%%%%%%%%%%%%%%%%%

\subsubsection{Analytical solution}
\label{analyticSolution}

Substituting $\hat{v}=\hat{r} f(\hat{z})$ in the differential equation \eqref{eqn_Bew_Gl}, with which the symmetry condition \eqref{RB1} is automatically satisfied and leads to 
\begin{equation}
    \pdv[2]{f(\hat{z})}{\hat{z}} = 0.
    \label{ODE_f}
\end{equation}
The boundary conditions \eqref{RB2_v1} and \eqref{RB3_v1} simplify to
\begin{equation}
    - \frac{\lambda_u}{h} \pdv{\hat{f}(\hat{z})}{\hat{z}} \bigg|_{\hat{z} = 0} + \hat{f}(0) = 0
    \label{RB2_v2}
\end{equation}
and
\begin{equation}\label{RB3_v2}
      \frac{\lambda_o}{h} \pdv{\hat{f}(\hat{z})}{\hat{z}} \bigg|_{\hat{z}=1} + \hat{f}(1)=1.
\end{equation}
The integration of the differential equation \eqref{ODE_f} considering the boundary conditions \eqref{RB2_v2} and \eqref{RB3_v2} leads to the velocity profile
\begin{equation}
    \hat{v}(\hat{r},\hat{z}) = \frac{\hat{r}}{1 + (\lambda_o + \lambda_u)/h}\left(\hat{z} + \frac{\lambda_u}{h} \right).
\end{equation}
Since the fluid is Newtonian, the shear stress $\tau_{\phi z}$ \cite{Spurk} can be expressed as 
\begin{equation}
    \tau_{\phi z} = \mu \pdv{v_{\phi}}{z} = \frac{\mu \Omega r}{h + \lambda_u + \lambda_o}.
\end{equation}
By integrating the shear stress $\tau_{\phi z}$ over the area of the bottom plate, the torque becomes 
\begin{equation}
    M = \int_{\phi = 0}^{\phi = 2\pi} \int_{r=0}^{r=R} r\tau_{\phi z} r \, \mathrm{d}r \, \mathrm{d}\phi = \frac{\mu I_p \Omega}{h+\lambda_u+\lambda_o},
    \label{momentcalculate}
\end{equation}
where $I_p = {\pi}R^4/2$ represents the polar moment of the area of the lower plate.

For this study, we consider a homogeneous material for both rheometer plates and, hence, the slip length is assumed to be equal for both plates. In this case, the formula for the torque simplifies to equation \eqref{eqn:torque-formula}.
%%%%%%%%%%%%%%%%%%%%%%%%%%%%%%%%%%%%%%%%%%%%%%%%%%%%%%%%%%%%%%

\subsection{Simulation model}

\begin{figure}[ht!]
    \includegraphics[width=8cm]{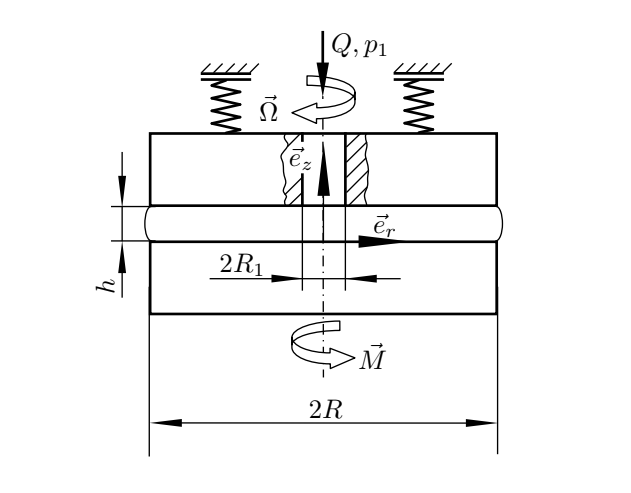} 
    \centering
    \caption{Principle sketch of the simulation model \cite{DissCorneli}.}
    \label{fig:sketch_Model2}
\end{figure}

\begin{figure}[ht!]
    \centering
    \captionsetup{position=top}
    \def\svgwidth{0.5\textwidth}
    {\footnotesize
        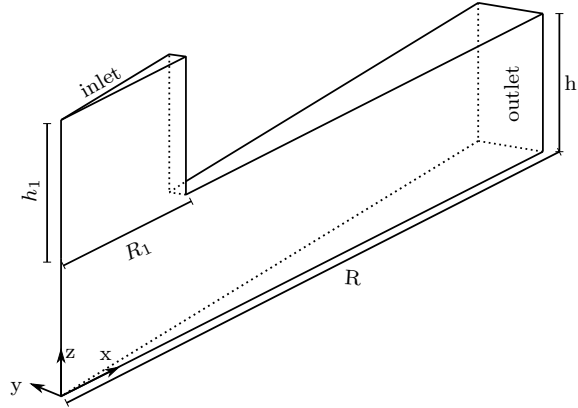
    }
    \vspace{0.5em}
    \caption{Schematic diagram of an axisymmetric domain with tribometer's radius $R$, gap height $h$ with an inlet height $h_1$ and inlet radius $R_1$. The inlet height $h_1$ vanishes for the \textit{simplified} simulation model.}
    \label{fig:schematic-diagram}
\end{figure}

% SLT is designed as sa i which evaluates the accuracy of the analytical model on which the measurement method is based.\todo{Isn't the measurement based on a simpler model?} 
Figure \ref{fig:sketch_Model2} shows the schematic of the SLT, simulated using the open-source software OpenFOAM~\cite{openfoam1,Maric2014}.   %This section presents the equations of motion applied in the computational domain and the corresponding boundary conditions.\todo{No substance}
%
%
%Compared with the analytical model, 
Two simulation models are presented: a \textit{simplified} and an \textit{actual} model. The schematic diagram of the axisymmetric computational domain is shown in Figure~\ref{fig:schematic-diagram}. In the \textit{simplified} model $(h_1 = 0)$, the cylindrical tube that drives the fluid between the plates is modeled as a hole (radius $R_1$), through which the fluid (density $\rho$, dynamic viscosity $\mu$) flows in with constant volume flow $Q$ at pressure $p_i$. The \textit{actual} simulation model discretizes the tube of height $h_1$ to estimate the effect of the inlet height on the resulting torque. At the outlet, the fluid leaves the computational domain. The simulation utilizes the OpenFOAM implementation of Navier slip~\citep{grundingNS} at the solid boundary.

%%%%%%%%%%%%%%%%%%%%%%%%%%%%%%%%%%%%%%%%%%%%%%%%%%%%%%%%%%%%%%%%

\subsubsection{Equations of motion}

Since OpenFOAM does not support the formulation of the balance equations in cylindrical coordinates, we consider the Navier-Stokes equations in Cartesian coordinates. Furthermore, we exploit the rotational symmetry of the problem to save computational effort by simulating only a section of the domain, as shown in Figure~\ref{fig:schematic-diagram}, using the axis-symmetric wedge boundary condition~\citep{openfoam1}. %\todo{CITE, see TODO in LaTex- yes, in Openfoam1, it is described, and the user guide is cited now.}

%TODO: For wedge, find if it is documented in
% \citep{openfoam1} - User Guide, if so, cite. 
% It is documented here:
% https://doc.cfd.direct/notes/cfd-general-principles/axisymmetric-wedge-condition
% Add a new citation for this page 
% as \miscellaneous and cite it, with \citep{}
% if you don't find wedge description 
% in the OpenFOAM.com User Guide
% or another CFD book. 

For the velocity vector $\boldsymbol{v}=[v_x, v_y, v_z]^T$ and the pressure $p$, the following balance equations for mass and momentum \cite{Spurk} apply in the computational domain
\begin{equation}
    \pdv{v_x}{x}+\pdv{v_y}{y}+\pdv{v_z}{z}=0,
\end{equation}
\begin{equation}
    \begin{aligned}
        & \rho \left( v_x \pdv{v_x}{x}+ v_y \pdv{v_x}{y} + v_z \pdv{v_x}{z} \right)=  -\pdv{p}{x} + \mu \left( \pdv[2]{v_x}{x}+\pdv[2]{v_x}{y}+\pdv[2]{v_x}{z} \right), \\
        & \rho \left( v_x \pdv{v_y}{x}+ v_y \pdv{v_y}{y} + v_z \pdv{v_y}{z} \right)=  -\pdv{p}{y} + \mu \left( \pdv[2]{v_y}{x}+\pdv[2]{v_y}{y}+\pdv[2]{v_y}{z} \right), \\
        & \rho \left( v_x \pdv{v_z}{x}+ v_y \pdv{v_z}{y} + v_z \pdv{v_z}{z} \right)=  -\pdv{p}{z} + \mu \left( \pdv[2]{v_z}{x}+\pdv[2]{v_z}{y}+\pdv[2]{v_z}{z} \right). \\
    \end{aligned}
\end{equation}

%%%%%%%%%%%%%%%%%%%%%%%%%%%%%%%%%%%%%%%%%%%%%%%%%%%%%%%%%%%%%%%%

\subsubsection{Boundary conditions}

The Navier slip boundary condition and the condition of impermeability hold at the bottom and top plates and the inlet channel wall, and they can then be expressed in the form
\begin{equation}
    - \lambda_u \left(\begin{array}{c}  \pdv{v_x}{z} + \pdv{v_z}{x} \\ \pdv{v_y}{z} + \pdv{v_z}{y} \\ 0 \end{array}\right)_{z=0} + \left(\begin{array}{c}  v_x \\ v_y \\ v_z  \end{array}\right)_{z=0}= \left(\begin{array}{c}  0 \\ 0 \\ 0\end{array}\right),
\end{equation}
\begin{equation}
     \lambda_o \left(\begin{array}{c}  \pdv{v_x}{z} + \pdv{v_z}{x} \\ \pdv{v_y}{z} + \pdv{v_z}{y} \\ 0 \end{array}\right)_{z=h} + \left(\begin{array}{c}  v_x \\ v_y \\ v_z  \end{array}\right)_{z=h}= \left(\begin{array}{c}  0 \\ r \Omega \\ 0\end{array}\right)_{z=h},
\end{equation}
\begin{equation}
     - \lambda_i \left(\begin{array}{c} 0 \\  \pdv{v_x}{y} + \pdv{v_y}{x} \\ \pdv{v_x}{z} + \pdv{v_z}{x}  \end{array}\right)_{x=R_1} + \left(\begin{array}{c}  v_x \\ v_y \\ v_z  \end{array}\right)_{x=R_1}= \left(\begin{array}{c}  0 \\ 0 \\ 0\end{array}\right)_{x=R_1},
\end{equation}
where $r=\sqrt{x^2+y^2}$ describes the distance between any point P on the upper plate with coordinates $(x,y,h)$ and the axis of rotation of the SLT.
At the inlet, the constant pressure $p_i$ boundary condition is applied, i.e., \
\begin{equation}
    p(x,y,h+h_1) = p_i \quad \text{for} \quad 0 \leq \sqrt{x^2+y^2} \leq R_1 %\left.\right|_{0 \leq r \leq R_1,z=h+h_1},
\end{equation}
where $p_i$ is obtained as a function of plate height $h$~\cite{DissCorneli} as
\begin{equation}\label{eqn:gapheight}
    h = \frac{\pi}{2k} \frac{R^2 - R_1^2}{\ln{\left(R/R_1\right)}} p_i.
\end{equation}
At the outlet, the ambient pressure is set to zero, i.e.\
\begin{equation}
    p(x,y,z) = 0 \quad \text{for} \quad \sqrt{x^2+y^2} = R \quad \text{and} \quad 0 \leq z \leq h. %\left.\right|_{r=R,0 \leq z \leq h} = 0,
\end{equation}
Furthermore, periodic boundary conditions are used on the sides of the computational domain to account for the rotational symmetry of the flow. 

%%%%%%%%%%%%%%%%%%%%%%%%%%%%%%%%%%%%%%%%%%%%%%%%

\subsubsection{Simulation setup}

The solution to the presented problem is obtained numerically using the unstructured Finite Volume Method and OpenFOAM ~\cite{openfoam1,Maric2014}, namely the \texttt{simpleFoam} solver for steady-state single-phase Navier-Stokes equations that utilizes the SIMPLE algorithm~\cite{openfoam1,simpleAlgo}. The input data of the simulation setup is publicly available, as well as the post-processing scripts and the secondary data \cite{SLTdatalib,dataSet}. Due to rotational symmetry, the computational domain is discretized using a prismatic (wedge) discretization in radial and axial directions. $N_r$ and $N_z$ denote the number of cells in radial and axial directions, respectively.
\begin{figure}[ht!]
    \includegraphics[width=10cm]{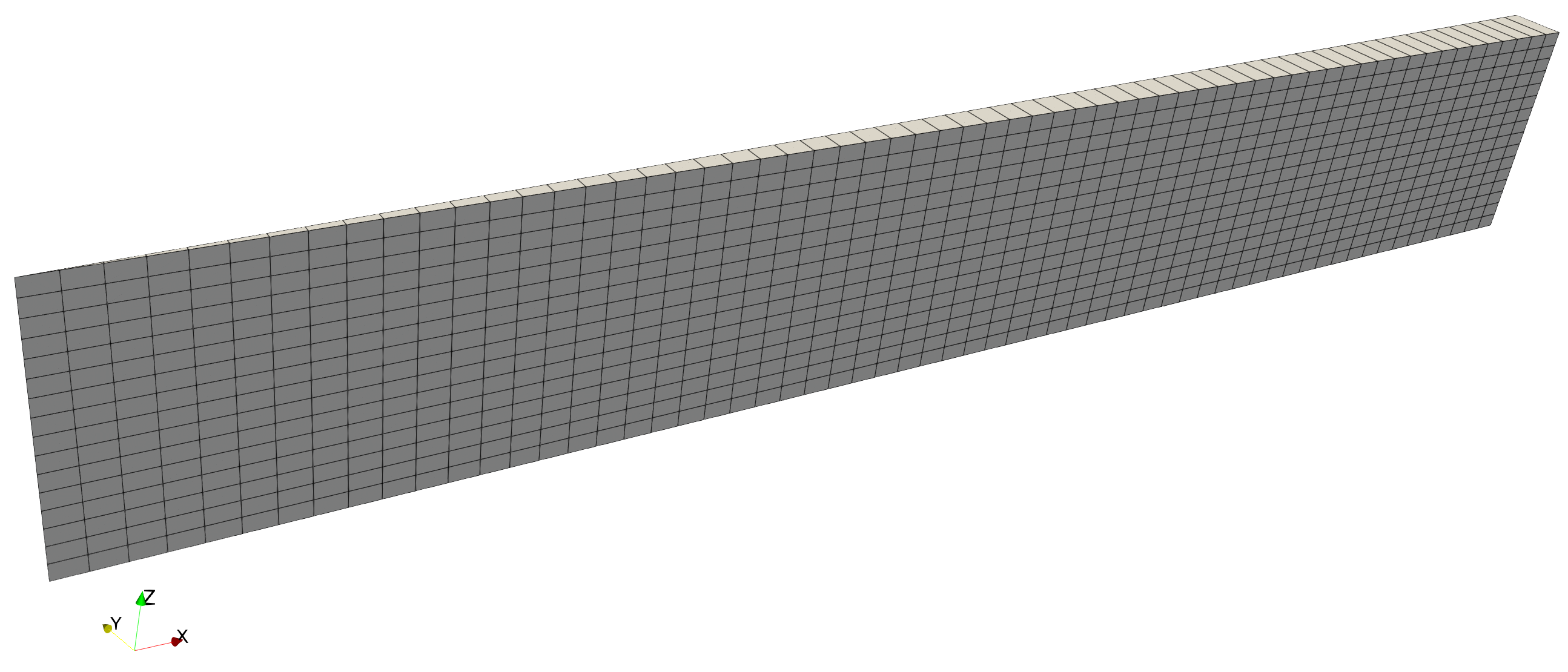} 
    \centering
    \caption{Discretized computational domain. The image is scaled in z-direction to improve the visibility.}
    \label{MeshModel2}
\end{figure}

%%%%%%%%%%%%%%%%%%%%%%%%%%%%%%%%%%%%%%%%%%

%% file: figures/analyticalModel.pdf_tex
%% Creator: Inkscape inkscape 0.92.5, www.inkscape.org
%% Accompanies image file 'analyticalModel.pdf' (pdf, eps, ps)
\begingroup%
  \makeatletter%
  \providecommand\color[2][]{%
    \errmessage{(Inkscape) Color is used for the text in Inkscape, but the package 'color.sty' is not loaded}%
    \renewcommand\color[2][]{}%
  }%
  \providecommand\transparent[1]{%
    \errmessage{(Inkscape) Transparency is used (non-zero) for the text in Inkscape, but the package 'transparent.sty' is not loaded}%
    \renewcommand\transparent[1]{}%
  }%
  \providecommand\rotatebox[2]{#2}%
  \newcommand*\fsize{\dimexpr\f@size pt\relax}%
  \newcommand*\lineheight[1]{\fontsize{\fsize}{#1\fsize}\selectfont}%
  \ifx\svgwidth\undefined%
    \setlength{\unitlength}{364.19034536bp}%
    \ifx\svgscale\undefined%
      \relax%
    \else%
      \setlength{\unitlength}{\unitlength * \real{\svgscale}}%
    \fi%
  \else%
    \setlength{\unitlength}{\svgwidth}%
  \fi%
  \global\let\svgwidth\undefined%
  \global\let\svgscale\undefined%
  \makeatother%
  \begin{picture}(1,0.51706443)%
    \lineheight{1}%
    \setlength\tabcolsep{0pt}%
    \put(0,0){\includegraphics[width=\unitlength,page=1]{analyticalModel.pdf}}%
    \put(0.49014592,0.23670661){\color[rgb]{0,0,0}\makebox(0,0)[lt]{\lineheight{1.25}\smash{\begin{tabular}[t]{l}$r$\end{tabular}}}}%
    \put(0.40737187,0.30159415){\color[rgb]{0,0,0}\makebox(0,0)[lt]{\lineheight{1.25}\smash{\begin{tabular}[t]{l}$z$\end{tabular}}}}%
    \put(0.31865396,0.4571406){\color[rgb]{0,0,0}\makebox(0,0)[lt]{\lineheight{1.25}\smash{\begin{tabular}[t]{l}$\phi, \Omega$\end{tabular}}}}%
    \put(0.48044598,0.08945278){\color[rgb]{0,0,0}\makebox(0,0)[lt]{\lineheight{1.25}\smash{\begin{tabular}[t]{l}$M$\end{tabular}}}}%
    \put(0,0){\includegraphics[width=\unitlength,page=2]{analyticalModel.pdf}}%
    \put(0.93992453,0.21833762){\color[rgb]{0,0,0}\makebox(0,0)[lt]{\lineheight{1.25}\smash{\begin{tabular}[t]{l}$h$\end{tabular}}}}%
    \put(0,0){\includegraphics[width=\unitlength,page=3]{analyticalModel.pdf}}%
    \put(0.41423212,0.03705628){\color[rgb]{0,0,0}\makebox(0,0)[lt]{\lineheight{1.25}\smash{\begin{tabular}[t]{l}$2R$\end{tabular}}}}%
    \put(0,0){\includegraphics[width=\unitlength,page=4]{analyticalModel.pdf}}%
  \end{picture}%
\endgroup%

%% file: figures/domain.pdf_tex
%% Creator: Inkscape inkscape 0.92.5, www.inkscape.org
%% Accompanies image file 'domain.pdf' (pdf, eps, ps)
\begingroup%
  \makeatletter%
  \providecommand\color[2][]{%
    \errmessage{(Inkscape) Color is used for the text in Inkscape, but the package 'color.sty' is not loaded}%
    \renewcommand\color[2][]{}%
  }%
  \providecommand\transparent[1]{%
    \errmessage{(Inkscape) Transparency is used (non-zero) for the text in Inkscape, but the package 'transparent.sty' is not loaded}%
    \renewcommand\transparent[1]{}%
  }%
  \providecommand\rotatebox[2]{#2}%
  \newcommand*\fsize{\dimexpr\f@size pt\relax}%
  \newcommand*\lineheight[1]{\fontsize{\fsize}{#1\fsize}\selectfont}%
  \ifx\svgwidth\undefined%
    \setlength{\unitlength}{480.2843165bp}%
    \ifx\svgscale\undefined%
      \relax%
    \else%
      \setlength{\unitlength}{\unitlength * \real{\svgscale}}%
    \fi%
  \else%
    \setlength{\unitlength}{\svgwidth}%
  \fi%
  \global\let\svgwidth\undefined%
  \global\let\svgscale\undefined%
  \makeatother%
  \begin{picture}(1,0.71264559)%
    \lineheight{1}%
    \setlength\tabcolsep{0pt}%
    \put(0,0){\includegraphics[width=\unitlength,page=1]{domain.pdf}}%
    \put(0.12879583,0.54402494){\color[rgb]{0,0,0}\rotatebox{29.595051}{\makebox(0,0)[lt]{\lineheight{1.25}\smash{\begin{tabular}[t]{l}inlet\end{tabular}}}}}%
    \put(0.89203468,0.50063979){\color[rgb]{0,0,0}\rotatebox{89.893493}{\makebox(0,0)[lt]{\lineheight{1.25}\smash{\begin{tabular}[t]{l}outlet\end{tabular}}}}}%
    \put(0,0){\includegraphics[width=\unitlength,page=2]{domain.pdf}}%
    \put(0.15497878,0.08153297){\color[rgb]{0,0,0}\makebox(0,0)[lt]{\lineheight{1.25}\smash{\begin{tabular}[t]{l}x\end{tabular}}}}%
    \put(0.09474661,0.08599461){\color[rgb]{0,0,0}\makebox(0,0)[lt]{\lineheight{1.25}\smash{\begin{tabular}[t]{l}z\end{tabular}}}}%
    \put(-0.00173644,0.02743545){\color[rgb]{0,0,0}\makebox(0,0)[lt]{\lineheight{1.25}\smash{\begin{tabular}[t]{l}y\end{tabular}}}}%
    \put(0,0){\includegraphics[width=\unitlength,page=3]{domain.pdf}}%
    \put(0.97257484,0.55725557){\color[rgb]{0,0,0}\makebox(0,0)[lt]{\lineheight{1.25}\smash{\begin{tabular}[t]{l}h\end{tabular}}}}%
    \put(0.58523597,0.21528893){\color[rgb]{0,0,0}\rotatebox{-3.937939}{\makebox(0,0)[lt]{\lineheight{1.25}\smash{\begin{tabular}[t]{l}R\end{tabular}}}}}%
    \put(0.20471582,0.25720937){\color[rgb]{0,0,0}\rotatebox{25.856449}{\makebox(0,0)[lt]{\lineheight{1.25}\smash{\begin{tabular}[t]{l}$R_1$\end{tabular}}}}}%
    \put(0,0){\includegraphics[width=\unitlength,page=4]{domain.pdf}}%
    \put(0.04361117,0.35974186){\color[rgb]{0,0,0}\rotatebox{90.099776}{\makebox(0,0)[lt]{\lineheight{1.25}\smash{\begin{tabular}[t]{l}$h_1$\end{tabular}}}}}%
  \end{picture}%
\endgroup%

%% file: sections/results.tex
\section{Results and Discussion}
\label{sec:results}

In this section, a convergence study is performed to evaluate the accuracy of the simulation model. Subsequently, the results of both modeling approaches are presented and compared with the measured data. The parameters used for the simulation studies are presented in Table \ref{TableParameterModel}. The slip lengths value $\lambda = \lambda_o = \lambda_u = 540 \text{nm}$ results from fitting experimental measurements to the analytical model \eqref{eqn:torque-formula} (see \cite{DissCorneli} for details). In the following, we aim to validate the analytical model with simulations. Since the inlet is composed of the same material, we will assume that $\lambda_i = \lambda =  540 \text{nm}$.
\begin{table}[h]
    \centering
        \begin{tabular}{|l|l|}
            \hline
            \textbf{Parameter}   &  \textbf{Value}        \\ \hline
            $R$                  &  32 mm                 \\ \hline
            $R_1$                &  1 mm                  \\ \hline
            $h$                  &  2 \dots 10 \textmu m  \\ \hline
            $h_1$                &  259 cm                \\ \hline
            $\Omega$             &  4$\pi$~rad/s          \\ \hline
            $\mu$                &  0.039 Pa.s            \\ \hline
            $\lambda_u$          &  540 nm                \\ \hline
            $\lambda_o$          &  540 nm                \\ \hline
            $\lambda_i$          &  540 nm                \\ \hline
        \end{tabular}
    \caption{Parameters of the simulation model.}
    \label{TableParameterModel}
\end{table}
%

%%%%%%%%%%%%%%%%%%%%%%%%%%%%%%%%%%%%%%%%%%%%%%%%%%%%%%%%%%%

\subsection{Mesh convergence study}

\begin{figure}[h]
\includegraphics[width=0.6\textwidth]{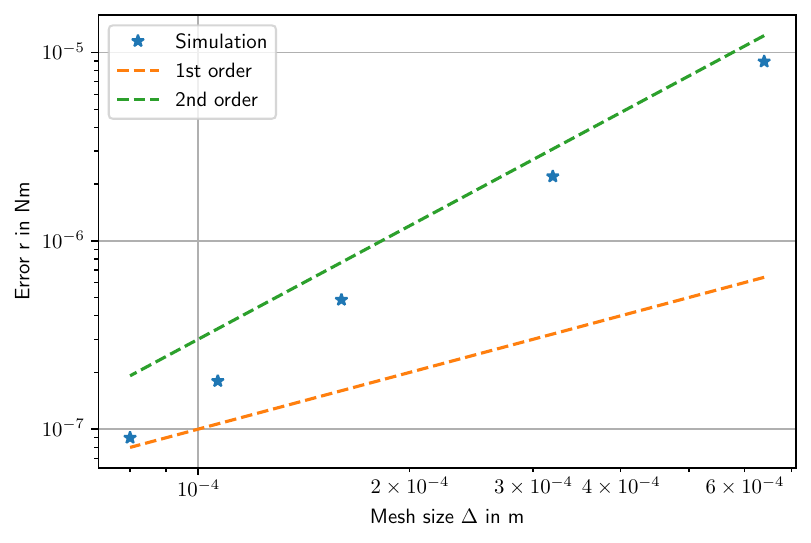} 
\centering
\caption{Convergence analysis for the (\textit{simplified}) simulation model.}
\label{convergenceStudy}
\end{figure}

In the following study, the torque on the bottom plate $M$ is calculated numerically for different mesh sizes, presented in Table \ref{GridsUsed}. The mesh width $\Delta$ is defined by averaging the mesh widths $\Delta_r$ and $\Delta_z$ in the radial and axial directions, respectively, and is given as
\begin{equation}
    \Delta = \sqrt{\Delta_r^2 + \Delta_z^2} 
           = \sqrt{\left(\frac{R}{N_r}\right)^2 + \left(\frac{h}{N_z}\right)^2}.
\end{equation}
Furthermore, the error $e$ in the torque calculations is defined as
\begin{equation}
    e=|M-M_{ref}|,
\end{equation}
where the reference torque $M_{ref}$ is the torque resulting from the simulation with the finest-resolution mesh M6 (see Table~\ref{GridsUsed}).
\begin{table}[h]
    \centering
    \begin{tabular}{|l|l|l|l|l|l|l|}
        \hline
        \textbf{Mesh}       & M1  & M2   & M3   & M4    & M5   & M6     \\ \hline
        $N_r$               & 50  & 100  & 200  & 300   & 400  & 1000   \\ \hline
        $N_z$               & 2   & 4    & 8    & 12    & 16   & 40     \\ \hline
        $\Delta[\mu m]$     & 640 & 320  & 160  & 107   & 80   & 32     \\ \hline
    \end{tabular}
    \caption{Parameters used in the mesh convergence analysis.}
    \label{GridsUsed}
\end{table}
The results of the convergence study are illustrated in Figure \ref{convergenceStudy}. As expected, the results demonstrate a second-order convergence rate for the error $e$, given that a second-order discretization method is employed to solve the Navier-Stokes equation. Additionally, the torque $M$ is computed as the integral of the local velocity field derivative over the area of the lower plate (see \cref{momentcalculate}). %Consequently, a second-order error is expected.
\subsection{Simulation results and validation}
This section compares the velocity profiles obtained through simulation models with those derived from the analytical model. Moreover, it includes a comparative analysis of the torque values computed from the analytical model, simulations, and experimental studies. This section aims to provide insights into the reliability of the models introduced in \cref{sec:methodsAndModels}. The results presented in this section are obtained using the mesh M6 (see \cref{GridsUsed}) and a tribometer plate gap height $h=5\mu m$.

\subsubsection{Velocity profiles}
\begin{figure}[h]
    \includegraphics[width=0.6\textwidth]{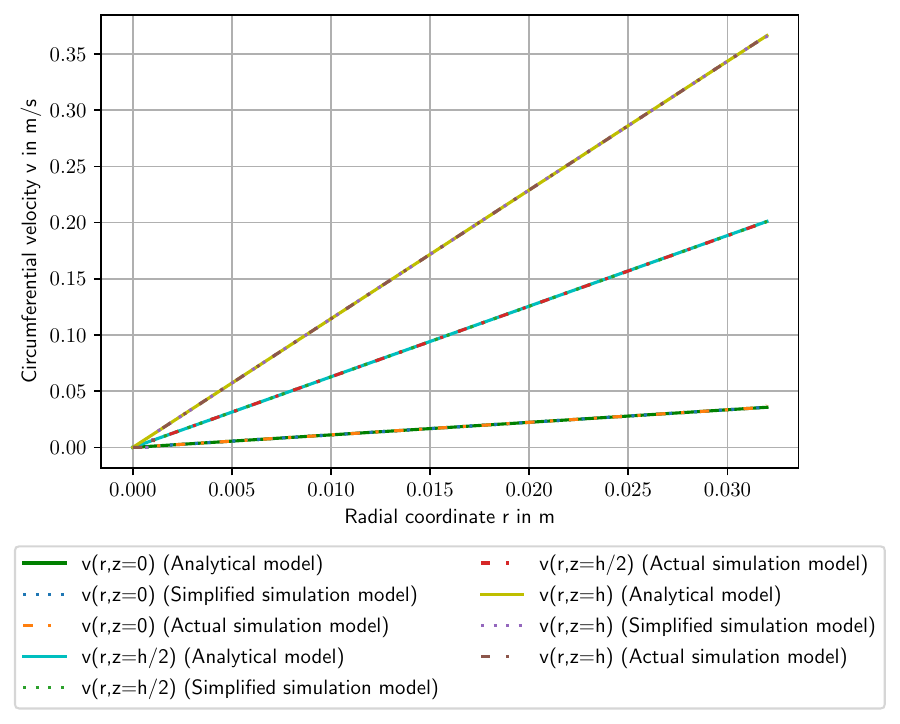} 
    \centering
    \caption{Comparison of the circumferential velocity $v$ along the radial direction.}
    \label{model1vs2Ur}
\end{figure}

\begin{figure}[h]
    \includegraphics[width=0.6\textwidth]{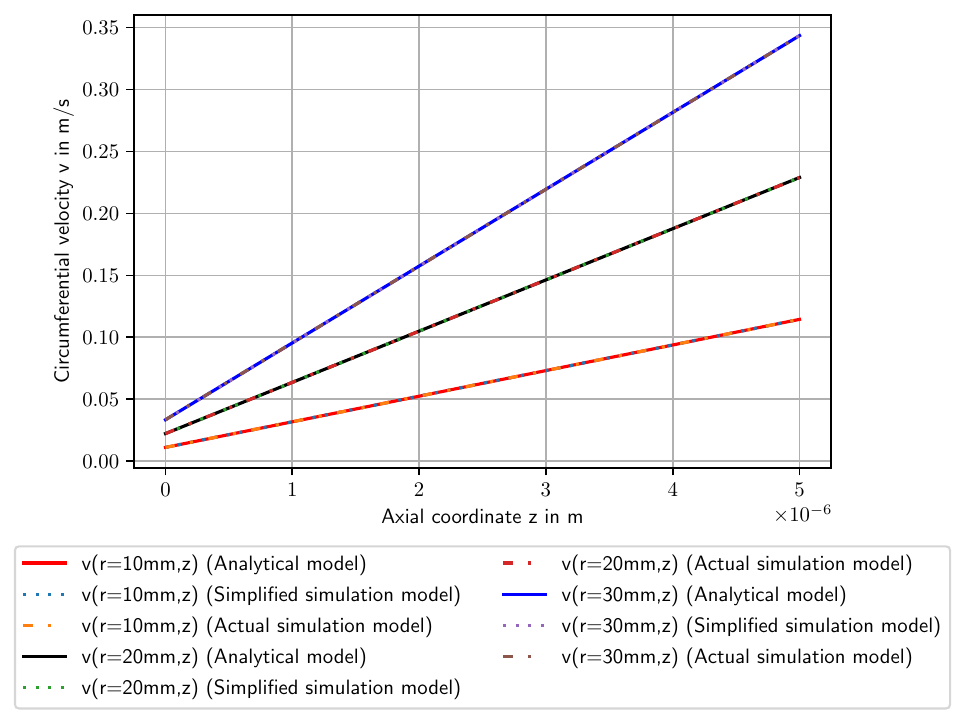} 
    \centering
    \caption{Comparison of the circumferential velocity $v$ along the axial direction.}
    \label{Model1vs2Uz}
\end{figure}

Figure \ref{model1vs2Ur} shows the velocity $v$ along the radial direction for three different $z$ values. Figure \ref{Model1vs2Uz} shows the dependence of velocity $v$ on axial coordinate $z$ for three different $r$ values. It can be seen that simulation results are in excellent agreement with the analytical model. It is also observed that the inlet height $h_1$ in the \textit{actual} simulation model does not affect the circumferential velocity component. 

\begin{figure}[h]
    \includegraphics[width=0.6\textwidth]{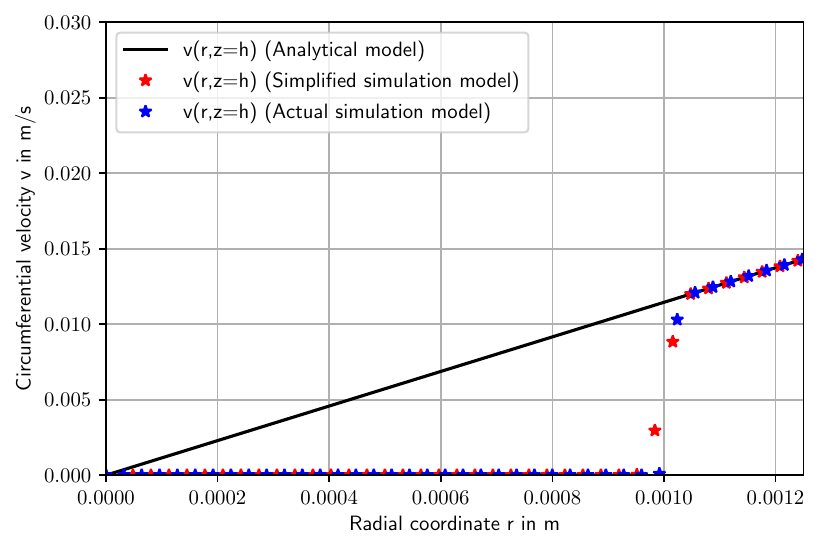} 
    \centering
    \caption{Comparison of the circumferential velocity $v$ along the radial direction in the inflow region. Even for an extremely large inner radius of the inlet pipe with $0$ axial velocity, the flow almost immediately recovers the analytical model for the axial velocity after exiting the inlet.}
    \label{Model1vs2UrZoomed}
\end{figure}

However, the analytical and simulation models differ in the inflow area, as depicted in Figure \ref{Model1vs2UrZoomed}. This deviation can be expected because the inflow is not modeled in the analytical model. Moreover, the deviation does not affect the torque at the bottom plate since the torque depends only on the partial derivative $\pdv{v}{z} \big|_{z=0}$. 
%

%%%%%%%%%%%%%%%%%%%%%%%%%%%%%%%%%%%%%%%%%%%%%%5
% \todo{In my opinion, the result that the inlet effect, i.e. the separation bubble at the edge, has no effect on the measuremnts and the model could be explained more. I would suggest having a magnified picture of the bubble in a countour plot of the pressure for example and calculate the length to show that its influence negligible due to the fact that the remaining surface area is so much larger.}
\subsubsection{Torque}

The available experimental measurements of torque $M$ and the tribometer plate gap height $h$, along with the operating parameters listed in Table  \ref{OperatingParameters}, are used to validate the models for torque measurement. The relationship of $M^{-1}$ and $h$ can be very well represented by the straight line equation $M^{-1}=bh+a$. The parameters $a$ and $b$ are determined by the Least Squares Method \cite{DissCorneli}.
\begin{table}[h]
    \centering
    \begin{tabular}{|l|l|}
        \hline
        \textbf{Parameter}   & \textbf{Value}         \\ \hline
        $h$                  &  2 \dots 10 \textmu m  \\ \hline
        $\Omega$             &  8$\pi$~rad/s          \\ \hline
    \end{tabular}
    \caption{Operating parameters of the SLT used to validate the models.}
    \label{OperatingParameters}
\end{table} 

The analytical model measures the slip lengths of the upper and lower tribometer's plates, $\lambda_u$ and $\lambda_o$, respectively, and the dynamic viscosity $\mu$ (see \cref{momentcalculate}). Subsequently, the following applies
\begin{equation}
    \begin{aligned}
        &\lambda_u=\lambda_o=\frac{a}{2b}, \\
        &\mu=\frac{1}{b \Omega I_p}.
    \end{aligned}
\end{equation}
Furthermore, these physical parameters are then used to evaluate the analytical and simulation models.

The measurements of the tribometer plate gap height $h$ and the moment $M$ are subject to errors. In addition, temperature changes in the lubrication gap or variations in motor frequency can distort the measurements. The error propagation of the mentioned errors results in the uncertainties $\delta a$ and $\delta b$ of the straight-line equation parameters. From this, the $95\%$ confidence interval is derived, within which the measuring points are located with a probability of $95\%$.

\begin{figure}[h]
    \includegraphics[width=0.6\textwidth]{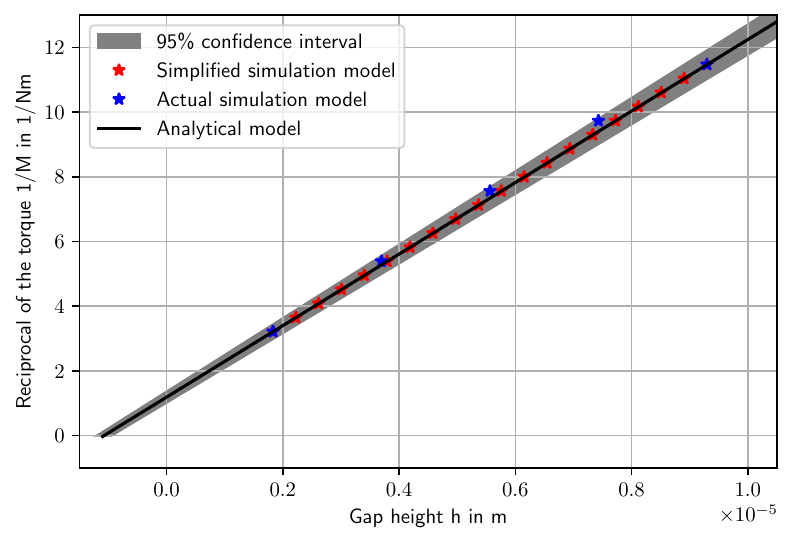} 
    \centering
    \caption{Validation of the models with the measurement of the torque $M$ as a function of the gap width $h$.}
    \label{ValidationM}
\end{figure}

Figure \ref{ValidationM} shows the $95\%$ confidence interval of the measured data and the corresponding results of the analytical and simulation models. It is noticeable that the results of both modeling approaches agree very well with the measurement data, showing that using the analytical model is justified for measuring the slip length.

%% file: sections/conculsion.tex
\section{Summary}
\label{sec:conclusion}
This work aims to verify and validate the simplified analytical model underlying the measurement principle of the Darmstädter Slip Length Tribometer (DSLT). The DSLT is a classical plate tribometer developed to measure the effect of surface roughness in the form of slip length of the Navier-slip boundary condition. 
The simulation results obtained by discretizing the full incompressible Navier-Stokes equations with the Navier-slip boundary condition in OpenFOAM are used to validate the measurements and analytical results. The results show very good agreement between the experimental measurements, the analytical model, and the simulation models, justifying the use of the simplified analytical model as the basis of the measurement principle of the DSLT.
%with the framework of the Collaborative Research Centre (CRC) 1194 \todo{SLT was independently developed at FST}